\documentclass[final,3p,times,twocolumn]{elsarticle}
 \biboptions{comma,sort&compress}
\usepackage{here}
 \usepackage{graphicx}
  \usepackage{epsfig}
\usepackage{comment}

\def\beq{\begin{equation}}
\def\eeq{\end{equation}}
\def\bea{\begin{eqnarray}}
\def\eea{\end{eqnarray}}

\def\Tr{{\rm Tr}}
\newcommand{\fm}{\,{\rm fm}}
\newcommand{\MeV}{\,{\rm MeV}}
\newcommand{\GeV}{\,{\rm GeV}}
\newcommand{\ignore}[1]{}

\journal{Nuc. Phys. (Proc. Suppl.)}

\begin{document}

\begin{frontmatter}

\title{Heavy ${\bar Q}Q$  free energy from hadronic states$^*$}
 \cortext[cor0]{Talk given at 18th International Conference in Quantum Chromodynamics (QCD 15,  30th anniversary),  29 June - 3 July 2015, Montpellier - FR.}
 \author[label1]{E.~Meg\'{\i}as\fnref{fn1}}
  \fntext[fn1]{Speaker, Corresponding author.}
  \ead{emegias@mppmu.mpg.de}
\address[label1]{Max-Planck-Institut f\"ur Physik (Werner-Heisenberg-Institut), F\"ohringer Ring 6, D-80805, Munich, Germany}

 \author[label2]{E.~Ruiz Arriola}
 \ead{earriola@ugr.es}
 
  \author[label2]{L.L.~Salcedo}
  \ead{salcedo@ugr.es}

\address[label2]{Departamento de F{\'\i}sica At\'omica, Molecular y Nuclear
  and Instituto Carlos I de F{\'\i}sica Te\'orica y Computacional, Universidad
  de Granada, E-18071 Granada, Spain.}

\pagestyle{myheadings}
\markright{ }
\begin{abstract}
Within the spirit of the Hadron Resonance Gas model, we study a representation
of the heavy $\bar{Q}Q$ free energy at temperatures below the phase transition
in terms of the string and heavy-light hadrons. We discuss the string breaking
phenomenon and the relevance of avoided crossings between the fundamental
string and the hadron spectrum. Good agreement with lattice data is achieved.
\end{abstract}

\end{frontmatter}

\section{Introduction}
\label{sec:introduction}

A fruitful approach to study the confined phase of QCD is the Hadron Resonance
Gas (HRG) model, in which the equation of state is obtained by assuming a gas
of non-interacting massive stable and point-like particles, which are taken as
the conventional hadrons~\cite{Hagedorn:1984hz}. This model has arbitrated the
discrepancies between the different collaborations in the lattice
community~\cite{Huovinen:2009yb}.

The effort to study the thermodynamics of QCD has led also to
important advances in the study of the interquark forces at finite
temperature. The vacuum expectation value of the Polyakov loop is
related to the propagator of a static quark, and it has been used as
an order parameter for the confinement/deconfinement transition of
color charges~\cite{Svetitsky:1985ye}. The free energy of a heavy
${\bar Q}Q$ pair in a thermal medium, which can be computed from the
correlation function of Polyakov loops, has important phenomenological
consequences in the study of quarkonia physics in heavy ion
experiments~\cite{Brambilla:2010cs}. In line with the HRG approach we
have proposed recently~\cite{Megias:2012kb,Megias:2013xaa} a hadronic
representation for the vacuum expectation value of the Polyakov loop
which has been quite successful in describing the available lattice
data up to $T \simeq 180 \MeV$~\cite{Megias:2012kb,Bazavov:2013yv}. In
this communication we represent the heavy ${\bar Q}Q$ free energy in
terms of the HRG, and compare with recent lattice simulations.

\section{Quark potential and string breaking}
\label{sec:quark_potential}

One of the most studied quantities in QCD is the static energy between
heavy sources, such as heavy quark systems~\cite{Necco:2001xg} and 
also for more exotic states like
gluelumps~\cite{Simonov:2000ky,Megias:2014bfa}. We will present in
this section the most relevant properties of this interaction which
will be used later in the study of the heavy $\bar{Q}Q$ free energy.

\subsection{Static energies in QCD}
\label{sec:static_energies}

Inspired by perturbation theory to second order~\cite{Anzai:2010td}
the static interaction between heavy sources $A$ and $B$ in QCD is
often modeled implementing Casimir scaling, 
\begin{equation}
V_{AB}(r) = \lambda_A \cdot \lambda_B 
\left[ \frac{\alpha_S}{r} - \kappa r \right] \,,
\end{equation}
where $\lambda$ is the SU($N_c$) color group generator corresponding
to the representation of the source. The Coulomb-like term accounts
for the perturbative behavior from one-gluon exchange at short
distances, while the confining linearly growing term accounts for the
energy of the string connecting the two heavy quarks. Casimir scaling
implies a relation between the fundamental $Q\bar{Q} \equiv {\mathbf
  3} \times \bar{\mathbf 3}$ and adjoint $GG \equiv {\mathbf 8} \times
{\mathbf 8}$ color sources given by $V_{Q{\bar Q}}(r) = \frac{9}{4}
V_{GG}(r)$, and holds on the lattice for $N_c=3$~\cite{Bali:2000un}
and for heavy sources in the fundamental representation $Q\bar{Q}
\equiv {\mathbf 3} \times \bar{\mathbf 3}$ at
$N_c=3,5,7$~\cite{Albertus:2015fca}. The $N_c$ independence of the
$V_{Q \bar Q }$ potential has also been established.  For $N_c=3$ one
gets
\begin{equation}
V_{Q{\bar Q}}(r) = \sigma r - \frac{4\alpha_S}{3r}  \,, \label{eq:VCornell}
\end{equation}
where $\alpha_S\simeq \pi/16$ is the effective coupling and $\sigma
\equiv 4 \kappa /3 \simeq (0.42\GeV)^2$ is the string tension.

\subsection{String breaking}

Using this potential, one can study the transition between a
$Q\bar{Q}$ system and a meson-antimeson system. When the separation
between the two heavy quarks increases, the energy of the system rises
linearly due to the string tension term. Eventually this system
becomes unstable at a critical distance~$r_c$ when a light $\bar{q}q$
is created from the vacuum, so that a heavy-light meson-antimeson
$M\bar{M} \equiv (Q\bar{q})(\bar{Q}q)$ channel opens.

Neglecting the meson-antimeson interaction, the
energies of these channels are
\begin{eqnarray}
E_{Q\bar{Q}}(r) &=& m_{Q} + m_{\bar Q} + V_{Q\bar{Q}}(r) \,,   \\
V_{(Q\bar{q})(q\bar{Q})}(r) &=& 
\Delta_{Q{\bar q}} + \Delta_{q{\bar Q}} \equiv 2\Delta \,,
\end{eqnarray}
where $\Delta_{Q{\bar q}} = \lim_{m_Q\to \infty} \left( M_{Q{\bar q}} - m_Q
\right) $ is the mass of the heavy-light meson (the mass of the heavy quark
itself $m_Q$ being subtracted). Using these formulas, one can estimate the
string breaking to be $r_c \simeq 4 M_0/\sigma \sim
1.2\fm$~\cite{Arriola:2014bfa,Arriola:2015gra}, where $M_0$ is the
constituent quark mass, which is in fair agreement with lattice
results~\cite{Bali:2005fu}.

More generally, the $Q\bar{Q}$ state can decay into any of the many excited
states of the meson spectrum after ${\bar q}q$ pair creation from the vacuum,
i.e. $V_{(Q\bar{q})(q\bar{Q})}^{(n,m)}(r) = \Delta_{Q{\bar q}}^{(n)} +
\Delta_{q{\bar Q}}^{(m)}$, or to the baryon spectrum, as long as they have the
same quantum numbers as the $Q\bar{Q}$ system.\footnote{The mechanism of
  decaying into baryons implies the creation of two pairs of light quarks
  $\bar{q}q$, leading to the formation of two heavy-light baryons with one
  heavy quark.}

\subsection{Avoided crossings}

\begin{figure}[t]
\begin{center}
\epsfig{figure=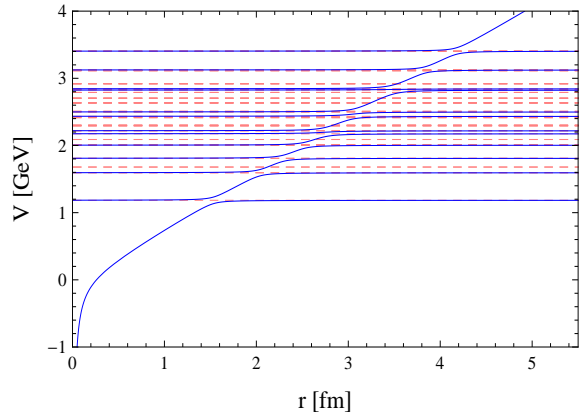,height=55mm,width=75mm}
\end{center}
\vspace{-0.5cm}
\caption{Adiabatic avoided crossing structure between the $Q{\bar Q}$
  fundamental state and the $M\bar{M}$ spectrum. We display as solid lines
  (blue) those states with avoided crossing, while dashed lines (red)
  stand for states with no avoided crossing, i.e. those with different quantum
  numbers than $Q\bar{Q}$. We have applied a mixing strength of
  $50\MeV$.}
\label{fig:avoided_crossing}
\end{figure}

In the picture presented above we have assumed a diabatic crossing
structure in which the states $Q\bar{Q}$ and $M\bar{M}$ are degenerate
at the point $r=r_c$ . However, the existence of a transition
potential $V_{Q\bar{Q} \to M\bar{M}}(r)$ between the two states,
implies the matrix interaction
\begin{equation}
V(r) = 
\left(
\begin{array}{cc}
V_{Q{\bar Q}}(r)  & V_{Q\bar{Q} \to M\bar{M}}(r)  \\
V_{Q\bar{Q} \to M\bar{M}}(r)  &  V_{M\bar{M}}(r) 
\end{array}
\right) \,.
\end{equation}
After diagonalization, the finite energy of the non-diagonal $Q\bar{Q}
\to M\bar{M}$ interaction lifts the degeneracy, a feature called level
repulsion, and there emerges the picture of adiabatic avoided crossing
shown in
Fig.~\ref{fig:avoided_crossing}~\cite{Arriola:2014bfa,Arriola:2015gra},
a phenomenon familiar from molecular physics in the
Born-Oppenheimer approximation~\cite{landau1965quantum}.

\section{Heavy $\bar{Q}Q$ free energy}
\label{sec:free_energy}

The free energy is (minus) the maximum work the system can perform at fixed
temperature. For a $\bar Q Q $ pair at separation~$r$ the
free energy is related to the correlator between Polyakov loops in the
following way \cite{McLerran:1980pk}
\begin{equation}
e^{-F_{Q\bar{Q}}(r,T)/T} =  
\langle \Tr_F \Omega (\vec r) \Tr_F \Omega(0)^\dagger \rangle 
 \,. \label{eq:Fave}
\end{equation}
This is the color averaged free energy, which follows from the thermal average
of singlet and adjoint free energies, see
e.g.~\cite{Jahn:2004qr}.\footnote{This normalization of the free energy differs by a factor $N_c^2$ from other conventions found in the literature.}
Several models have been proposed in the literature for the $\bar{Q}Q$ free
energy either in the confined~\cite{Andreev:2006nw,Colangelo:2010pe} or the
deconfined phase~\cite{Megias:2007pq,Riek:2010fk}. We focus on the former
case here.

\subsection{Hadronic representation}
\label{sec:hadronic_representation}

\begin{figure*}[htb]
\begin{tabular}{cc}
  \epsfig{figure=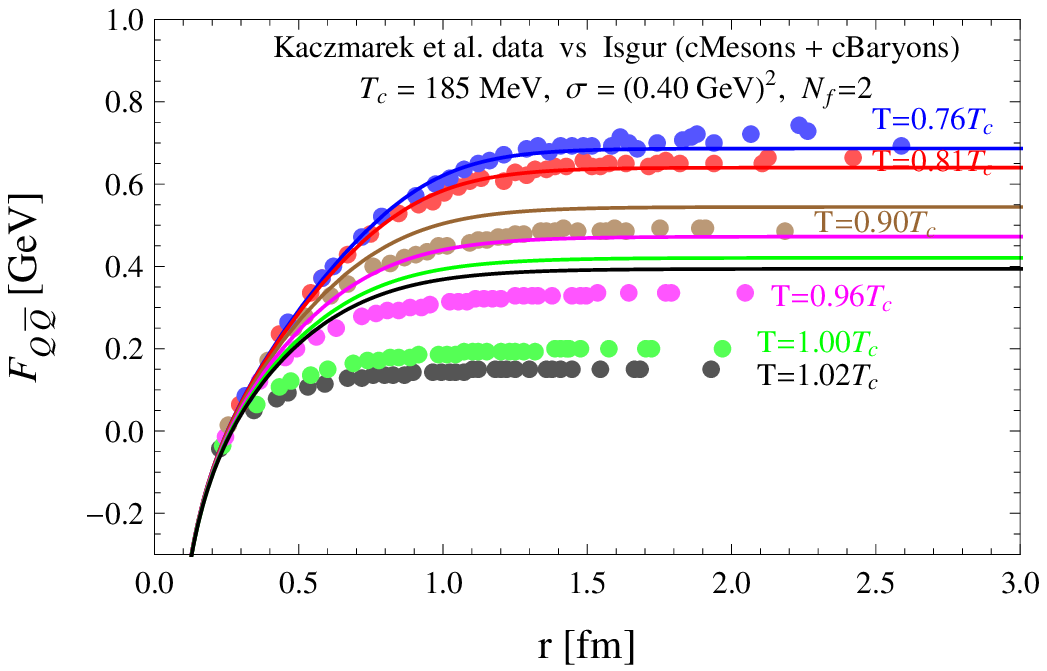,height=55mm,width=79mm} &
  \epsfig{figure=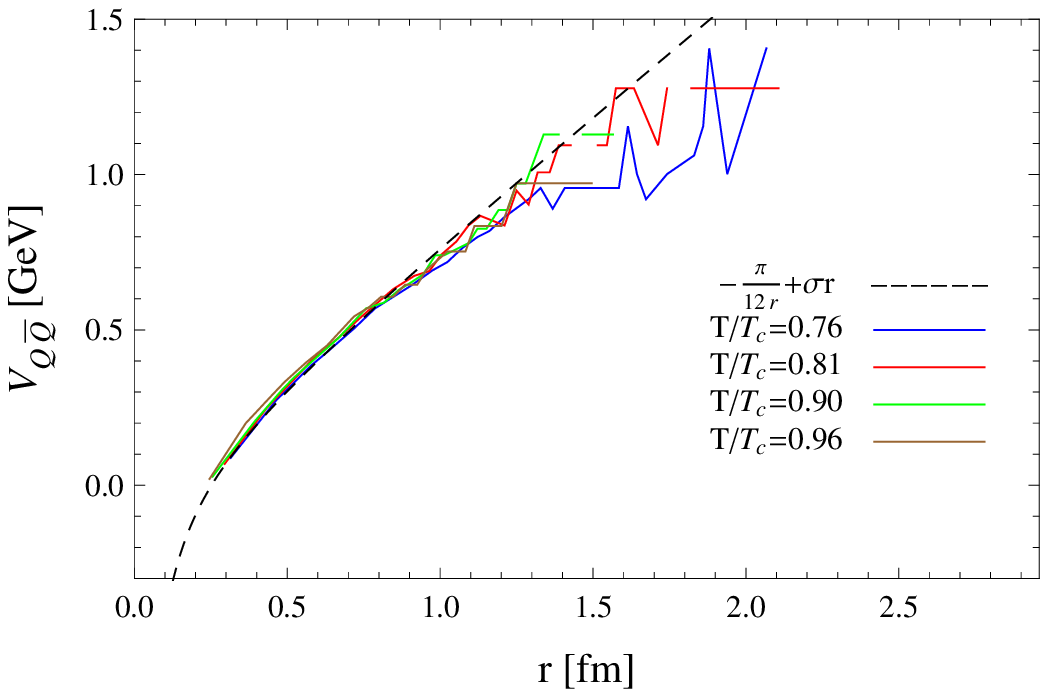,height=55mm,width=79mm} \\
(A)
       &
        (B)
\end{tabular}
\vspace{-0.4cm}
\caption{(A) Color averaged heavy ${\bar Q}Q$ free energy as a function of the
  separation. The dots are the lattice data for $N_f=2$ taken from
  Ref.~\cite{Kaczmarek:2005ui}. The solid lines represent the result by
  using Eq.~(\ref{eq:Fave2}) with the spectrum of heavy-light mesons and
  baryons with a charm quark, obtained with the Isgur model of
  Ref.~\cite{Godfrey:1985xj}. (B) Heavy $\bar{Q}Q$ potential at zero
  temperature as a function of the separation by using
  Eq.~(\ref{eq:VQQformula1}) and taking the lattice data shown in (A) for
  $T/T_c= 0.76,\, 0.81,\, 0.90$ and $0.96$. The dashed (black)
  line is the result of the Cornell potential, Eq.~(\ref{eq:VCornell}), with
  $\sigma=(0.40 \GeV)^2$.}
\label{fig:Fa076Isgur}
\end{figure*}

The correlation function in Eq.~(\ref{eq:Fave}) can be written as a ratio
of partition functions, with and without $Q\bar{Q}$ sources placed at a
distance $r$~\cite{Luscher:2002qv}. As argued in \cite{Arriola:2014bfa}, in
the confined phase such ratio admits an approximated spectral decomposition
\begin{equation}
e^{-F_{Q\bar{Q}}(r,T)/T} = e^{-V_{Q\bar{Q}}(r)/T} 
+ \sum_{n,m} e^{-V_{Q\bar{Q}}^{(n,m)}(r)/T}\,.
\end{equation}
By neglecting the avoided crossing, one finds
\begin{equation}
e^{-F_{Q\bar{Q}}(r,T)/T}  = e^{-V_{Q\bar{Q}}(r)/T} 
+ \left(\frac{1}{2}\sum_n e^{-\Delta_n /T} \right)^2 \,, \label{eq:Fave2}
\end{equation}
where $\Delta_n = \Delta^{(n)}_{q\bar{Q}} = \Delta^{(n)}_{\bar{q}Q}$, which
constitutes a hadronic representation of the heavy $\bar{Q}Q$ free
energy~\cite{Arriola:2014bfa,Arriola:2015gra}. In Fig.~\ref{fig:Fa076Isgur}(A)
we display the lattice results of \cite{Kaczmarek:2005ui} for the heavy
$\bar{Q}Q$ free energy along with the calculation from
Eq.~(\ref{eq:Fave2}). We use the spectrum of heavy-light mesons and baryons
with a charm quark and no strangeness, $N_f=2$, from the Isgur model
\cite{Godfrey:1985xj} up to $\Delta = 3.19\GeV$. We have taken $\sigma=(0.40
\GeV)^2$ and $\alpha_S = \pi/16$. We find an excellent agreement for
the lowest temperatures and for all distances.

The Polyakov loop expectation value can be computed from the large separation
limit of the correlator between Polyakov loops. From Eq.~(\ref{eq:Fave2}) one
has
\begin{equation}
L(T) := \lim_{r\to\infty}  e^{-F_{Q\bar{Q}}(r,T)/(2T)}  
=  \frac{1}{2}\sum_n e^{-\Delta_n /T}  \,. \label{eq:Lren}
\end{equation}
This is precisely the hadronic representation of the Polyakov loop introduced
in Ref.~\cite{Megias:2012kb}, which was shown to provide a good agreement with
lattice data for $T \lesssim 180\MeV$. This means that the spectrum from the
Isgur model already saturates the sum rules at these temperatures. It was
shown in~\cite{RuizArriola:2012wd} that Eq.~(\ref{eq:Lren}) can also be
obtained in chiral quark models coupled to the Polyakov, when one advocates
the local and quantum nature of the Polyakov
loop~\cite{Megias:2004hj,Megias:2006bn}.

A different exercise is to extract from the finite temperature lattice data
the zero temperature heavy $\bar{Q}Q$ potential. By using
Eqs.~(\ref{eq:Fave2}) and (\ref{eq:Lren}), one gets
\begin{equation}
V_{Q\bar{Q}}(r) = 
- T \log \bigg[ e^{-F_{Q\bar{Q}}(r,T)/T} - e^{-F_{Q\bar{Q}}(\infty,T)/T} \bigg] 
\,. 
\label{eq:VQQformula1}
\end{equation}
The terms in the argument of the $\log$-function are known from lattice data
and they are in general $T$-dependent, so that this is a nontrivial check.
The result for $V_{Q\bar{Q}}(r)$ is plotted in Fig.~\ref{fig:Fa076Isgur}(B).
This quantity turns out to be $T$-independent with a high accuracy, and it is
remarkably close to the $Q\bar{Q}$ Cornell potential, Eq.~(\ref{eq:VCornell}),
giving a strong indication of the validity of Eq.~(\ref{eq:Fave2}).

\begin{figure*}[t]
\begin{tabular}{cc}
  \epsfig{figure=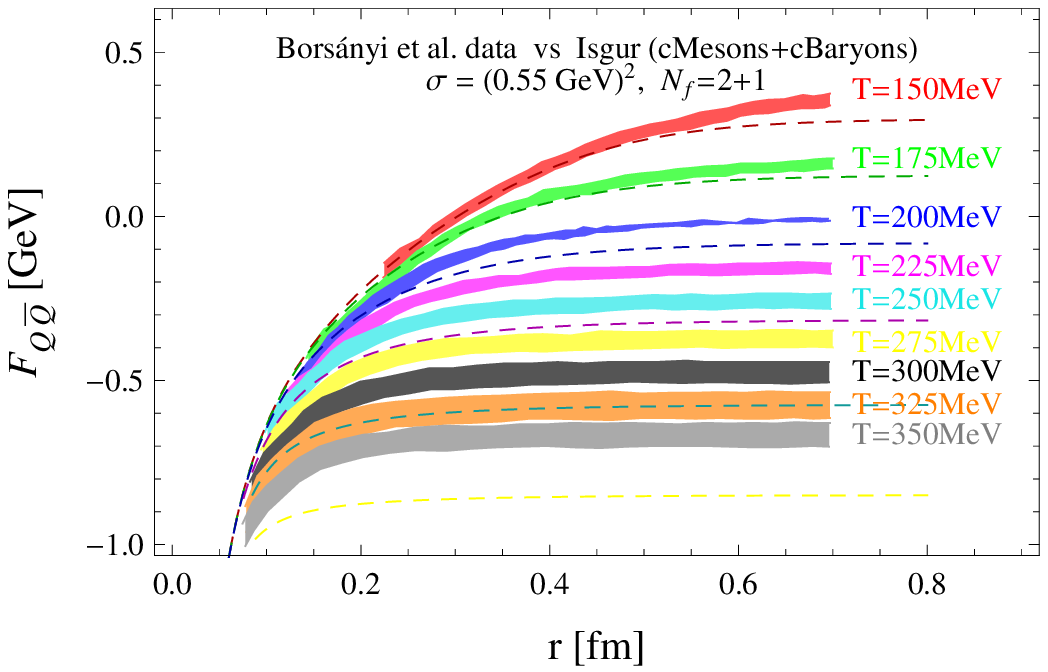,height=55mm,width=79mm} &
  \epsfig{figure=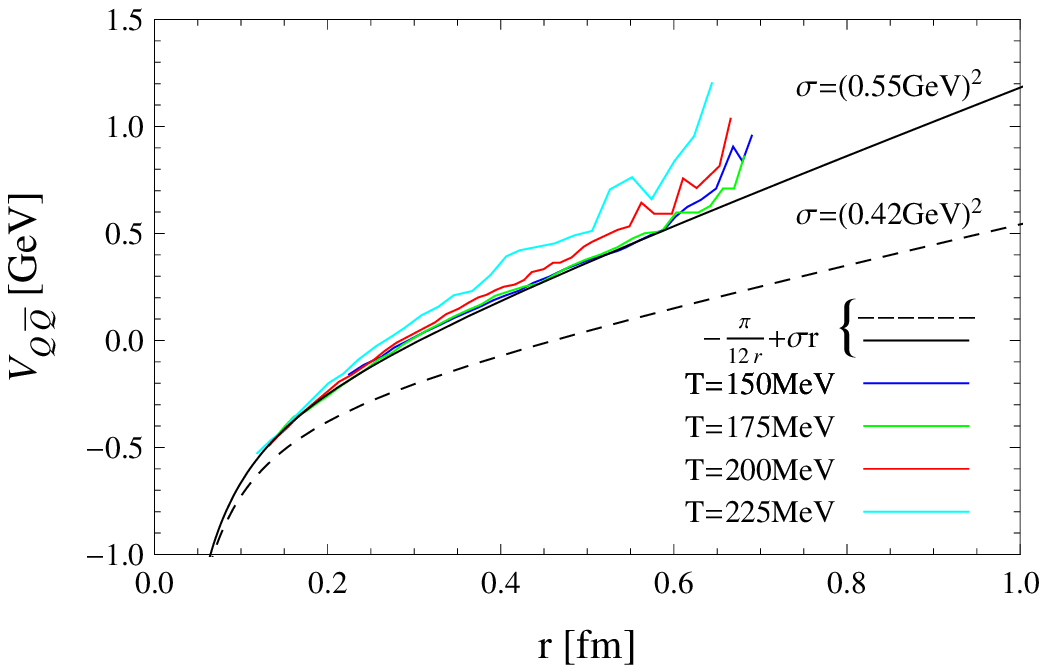,height=55mm,width=79mm} \\
(A)
       &
        (B)
\end{tabular}
\vspace{-0.4cm}
\caption{Same as Fig.~\ref{fig:Fa076Isgur} for the lattice data of
  Ref.~\cite{Borsanyi:2015yka}. We use $N_f=2+1$ for the meson and
  baryon spectrum and $\sigma=(0.55\GeV)^2$.}
\label{fig:FaZantowIsgur}
\end{figure*}

\subsection{Avoided crossing and heavy ${\bar Q}Q$ free energy}
\label{sec:avoided_crossing}

The lattice data of Ref.~\cite{Kaczmarek:2005ui} are computed for
2~light flavors with a pion mass of $m_\pi \simeq 770\MeV$, and
assumes the zero temperature ${\bar Q}Q$ potential behavior of the
free energy at short distances. The recent study of
Ref.~\cite{Borsanyi:2015yka} uses $N_f = 2+1$ with physical quark
masses, and releases the short distance assumption in the
renormalization procedure. The lattice results of
\cite{Borsanyi:2015yka} are displayed in
Fig.~\ref{fig:FaZantowIsgur}(A) along with our $N_f = 2+1$ calculation
assuming the hadronic representation of Eq.~(\ref{eq:Fave2}).  The
most striking observation is that a quite high value of the string
tension has to be assumed, $\sigma = (0.55\GeV)^2$, in order to
reproduce the lattice data of~\cite{Borsanyi:2015yka}, while $\sigma =
(0.40\GeV)^2$ worked for the data of~\cite{Kaczmarek:2005ui}. The same
conclusion is reached by extracting the zero temperature ${\bar Q}Q$
potential by using Eq.~(\ref{eq:VQQformula1}), as shown in
Fig.~\ref{fig:FaZantowIsgur}(B).

\begin{figure}[t]
\begin{center}
\epsfig{figure=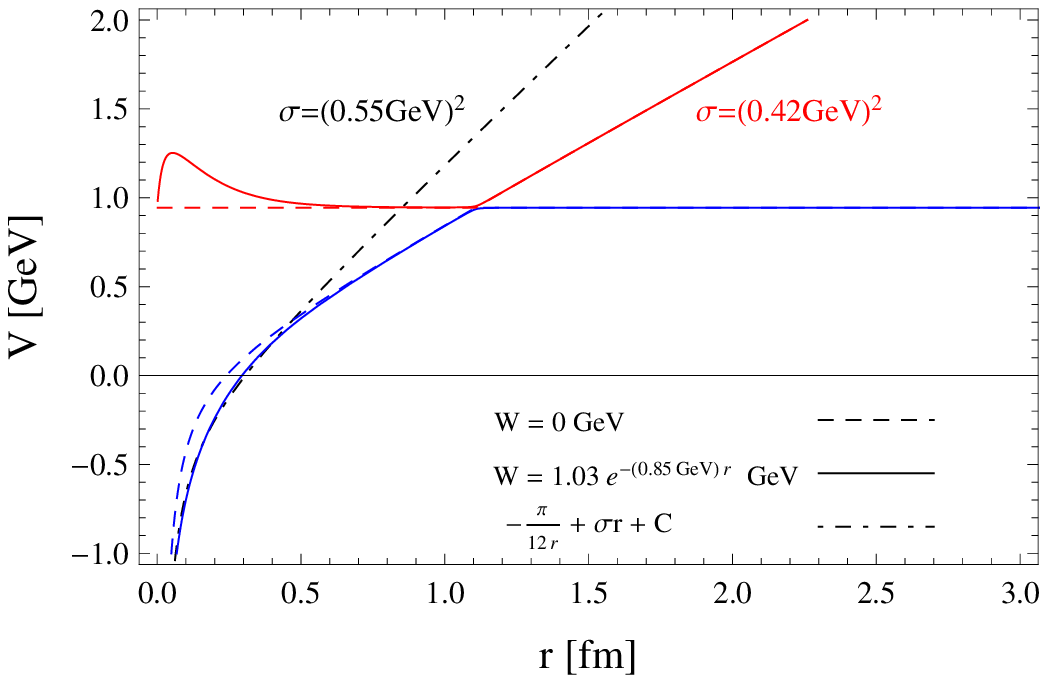,height=55mm,width=75mm}
\end{center}
\vspace{-0.5cm}
\caption{ With solid lines, string state and first meson-antimeson state as a
  function distance using the model of Eq.~(\ref{eq:V2states}) with $W(r) =
  W_0 \, e^{-mr}$. The parameters are obtained from a a best fit to the
  lattice data of the heavy $\bar{Q}Q$ free energy of
  Ref.~\cite{Borsanyi:2015yka} at the lowest available temperature, $T= 150
  \MeV$.  The dashed lines represent the result with this model
  taking $\sigma=(0.42\GeV)^2$ and no mixing, $W(r)=0$. The dot-dashed line
  (black) is the result from the Cornell potential, $V(r) =
  -\frac{\pi}{12r}+\sigma r + C$ with $\sigma = (0.55\GeV)^2$ and~$C
  = -0.3 \GeV$.}
\label{fig:plotFFodorT150}
\end{figure}

In view of this, one may wonder whether this effect in the string tension
could be effectively due to the existence of an appreciable mixing between the
string and the excited states in the lattice calculation
of~\cite{Borsanyi:2015yka}. To investigate this possibility, we take a simple
model consisting of only two coupled states, the ${\bar Q}Q$ pair (string
state) and the lightest meson-antimeson state:
\begin{equation}
V(r) = 
\left(
\begin{array}{cc}
-\frac{4\alpha_S}{3r} + \sigma r  & W(r)  \\
W(r)  &  2\Delta 
\end{array}
\right) \,. \label{eq:V2states}
\end{equation}
As a first estimate, we assume for the mixing a function of the form
\begin{equation}
W(r) = W_0 \, e^{-mr}  \,. \label{eq:Wexp}
\end{equation}
From a fit of the lowest temperature lattice data of ~\cite{Borsanyi:2015yka} for the heavy ${\bar Q}Q$ free energy, $T = 150\MeV$, one gets
\begin{eqnarray}
&&\hspace{-1cm}W_0 = 1.03(50) \GeV \,, \quad m = 0.85(38) \GeV \,, \nonumber \\
&&\hspace{-1cm}\sigma = (0.420(10)\GeV)^2 \,, \label{eq:fitFodorT150}
\end{eqnarray}
with $\chi^2/\mathrm{dof} = 0.036$. We have taken $2\Delta=0.944\GeV$ corresponding to the lowest heavy-light meson state of the
Isgur model. The result is displayed in Fig.~\ref{fig:plotFFodorT150} where
one can see that the mixing mimics a larger value of the string tension at
distances $r\lesssim 0.5 \fm$. The behavior of the first excited state at
short distances is in qualitative agreement with lattice simulations, see
e.g. Ref.~\cite{Bali:2005fu}. This result leads to the problem of the
identification of the parameter $m$ with the mass of some
particle. At first sight it would be tempting to assume that $m$ should be the
pion mass, however, the value we get for $m$ is heavier by a factor $\simeq
2\pi$.

\begin{figure}[t]
\begin{center}
\epsfig{figure=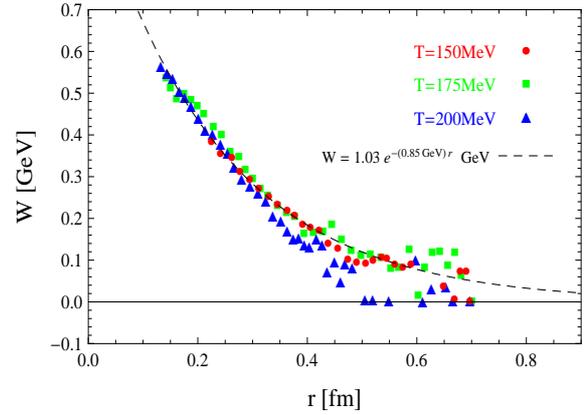,height=55mm,width=75mm}
\end{center}
\vspace{-0.5cm}
\caption{The transition potential $W$ of Eq.~(\ref{eq:V2states}) as a function
  of the $\bar{Q}Q$ separation. The dots reproduce the lattice data of the
  heavy ${\bar Q}Q$ free energy of Ref.~\cite{Borsanyi:2015yka} at different
  separations and temperatures assuming $\sigma = (0.42\GeV)^2$. The
  dashed line corresponds to the best fit obtained by assuming the functional
  form $W(r) = W_0 \, e^{-mr}$.}
\label{fig:plotW}
\end{figure}

The model dependence of this result can be relaxed by not assuming any
functional form for the mixing $W(r)$. To do so, for each point of the lattice
data \cite{Borsanyi:2015yka} at a given temperature we obtain the value of $W$
that allows the two states model, Eq.~(\ref{eq:V2states}), to reproduce the
free energy at that separation. The result for $W(r)$ computed in this way is
displayed in Fig.~\ref{fig:plotW}. A result quite similar to
Eqs.~(\ref{eq:Wexp})-(\ref{eq:fitFodorT150}) is obtained for $T \lesssim 200
\MeV$. The $T$-independence of $W(r)$ goes in favor of the reliability of the
model of Eq.~(\ref{eq:V2states}). For temperatures above $150\MeV$,
one would need to include a huge amount of states to reproduce the Polyakov
loop. Even after including all the available states of the Isgur model, the
sum rule for the Polyakov loop cannot be saturated for temperatures above
$180\MeV$~\cite{Megias:2012kb,Megias:2012hk}. For this reason, and
to avoid a source of error in $W(r)$ coming from the value of $F(\infty,T)$,
we have used for $F(\infty,T)$ the value extracted from the lattice data. This
is not necessary for the lowest temperature $T=150\MeV$, as the sum
rule is saturated in this case.

\section{Discussion and outlook}
\label{sec:discussion}

We have studied a natural extension of the HRG model to the heavy
$\bar{Q}Q$ free energy. This quantity has been computed in the
confined phase of QCD from a string and the HRG model with heavy-light
hadrons, and compared with existing lattice computations
\cite{Kaczmarek:2005ui,Borsanyi:2015yka} giving an excellent agreement
for $T\lesssim 200\MeV$. The relevance of string breaking at large
distances has been discussed as well. Finally, we have investigated
the possible role played by the avoided crossings between the string
and the spectrum of hadrons.

Our analysis of the avoided crossings leads to possible contradicting
results when analyzing the lattice data for the heavy ${\bar Q}Q$ free
energy from two different groups. While the interaction between the
string state and meson-antimeson states seems to be negligible in the
data from~\cite{Kaczmarek:2005ui}, a noticeable effect appears in the
analysis of the more recent data of~\cite{Borsanyi:2015yka}. This
could be well due to the different approaches for quark masses assumed
in these references, or the different renormalization
prescriptions. So, more accurate lattice data and better agreement
between different groups would be desirable on the one hand.  On the
other hand, our analysis can be improved either by including some
subleading effects like string excitations or extending the two states
mixing model with the inclusion of more states in the spectrum. Work
along these lines is in progress.

\section*{Acknowledgments}
Work supported by Spanish Ministerio de Econom{\'\i}a y Competitividad
and European FEDER funds under contract FIS2014-59386-P, by Junta de
Andaluc{\'\i}a grant FQM-225, and by the Spanish Excellence Network on
Hadronic Physics FIS2014-57026-REDT. The research of E.M. is supported
by the European Union (FP7-PEOPLE-2013-IEF) project
PIEF-GA-2013-623006.




\end{document}